# Epitaxial Growth and Electrical Properties of $VO_2$ on LSAT (111) substrate

Yang Liu, Shanyuan Niu, and Thomas Orvis,

Mork Family Department of Chemical Engineering and Material Science, University of Southern California, Los Angeles, CA 90089.

Haimeng Zhang

Ming Hsieh Department of Electrical Engineering, University of Southern California, Los Angeles, CA 90089.

Han Wang, and Jayakanth Ravichandran[a]

Mork Family Department of Chemical Engineering and Material Science, University of Southern California, Los Angeles, CA 90089

Ming Hsieh Department of Electrical Engineering, University of Southern California, Los Angeles, CA 90089.

[a] Electronic mail: jayakanr@usc.edu

We report the epitaxial growth and the electrical properties, especially the metal-to-insulator transition (MIT), of vanadium dioxide ($VO_2$) thin films synthesized on LSAT (111) ($[LaAlO_3]_{0.3}[Sr_2AlTaO_6]_{0.7}$) substrates by pulsed laser deposition. X-ray diffraction studies show that the epitaxial relationship between the $VO_2$ thin films and LSAT substrate is given as $VO_2(020)\|LSAT(111)$ and $VO_2[001]\|LSAT[11\bar{2}]$. We observed a sharp four orders of magnitude change in the longitudinal resistance for the $VO_2$ thin films around the transition temperature. We also measured distinct Raman spectra below and above the transition point indicating a concomitant structural transition between the insulator and metallic phases, in agreement with past investigations.

# I. INTRODUCTION

$VO_2$ is widely studied for its sharp metal-to-insulator transition (MIT) over a very narrow range of temperatures close to room temperature (~ 340 K).[1,2] At this transition temperature, $VO_2$ undergoes a large change in resistivity, as high as five orders of magnitude, in high-quality single crystals.[3] This transition is accompanied by a structural transition from monoclinic (M1) to tetragonal rutile (R) structure. The role of this structural change, and the strong correlation on the physical properties and MIT in $VO_2$ remains an important topic of investigation. The sudden and large resistivity change in $VO_2$ near room temperature has attracted tremendous attention for electronic and photonic applications such as logic devices,[4] oscillators,[5] filters,[6] thermal regulation,[7] and smart windows[8]. The MIT can also be induced by other impulses such as ultrafast lasers, where the transition can be achieved in sub-picosecond timescales,[9] making it an attractive material for ultrafast optical switches and sensors.

Thin film growth of $VO_2$ is of fundamental importance, as many device applications rely on this platform and one can achieve meta-stable strain-induced phases in thin films.[10] Thin films of $VO_2$ have been synthesized using various vapor phase deposition techniques such as chemical vapor deposition, reactive electron-beam evaporation, radio frequency sputtering and pulsed laser deposition (PLD).[11–17] As the vanadium-oxygen phase diagram[18] consists of principle oxides such as VO, $V_2O_3$, $VO_2$, $V_2O_5$, Magnéli phases with a general formula $V_nO_{2n-1}$ and composition between $VO_2$ and $V_2O_3$, and layered phases with a general formula $V_nO_{2n+1}$ and composition between $VO_2$ and $V_2O_5$, precise control of the oxygen stoichiometry is critical to achieving the desired phase. Hence, thin film growth of $VO_2$ is often sensitive to growth conditions such as oxygen partial pressure, which regulates the stoichiometry of the vanadium oxides. Pulsed laser deposition (PLD) is a powerful technique for the fabrication of high-quality VO$_2$ films as it allows precise control of the stoichiometry by tuning the deposition parameters.

Another important limitation to the growth of high-quality VO$_2$ films is the lack of single crystal substrates with a good epitaxial match. To maintain favorable conditions for hetero-epitaxy, VO$_2$ thin films have been grown on rutile TiO$_2$ substrates, which are isostructural to the high temperature phase of rutile VO$_2$.[19] The lattice parameters of rutile VO$_2$ are slightly smaller than rutile TiO$_2$, and this leads to small tensile strain for different orientations of rutile TiO$_2$ substrates.[20] Epitaxial growth of VO$_2$ thin films on substrates with 3$m$ symmetry, such as ZnO (0001) and Al$_2$O$_3$ (0001), has been investigated in the past. These substrates produce high-quality VO$_2$ thin films with three orders of magnitude change in resistance around the transition temperature, but structural domains along the in-plane direction in such films are randomly distributed and hard to control.[21,22] Cubic perovskite substrates with (111) orientation possess 3$m$ symmetry and can provide the desired epitaxial surface for the growth of VO$_2$ thin films. (111)-oriented SrTiO$_3$ (STO) has been successfully used as the substrate for VO$_2$ films with excellent electrical properties.[10] Despite these desirable properties, structural characterization of these films using X-ray diffraction of these films is challenging as the out-of-plane $d$-spacing for VO$_2$ (010) and STO (111) are close to each other. Moreover, the structure factor of VO$_2$ is much smaller than STO, further complicating structural analysis. [LaAlO$_3$]$_{0.3}$[Sr$_2$AlTaO$_6$]$_{0.7}$ (LSAT) is one of the widely used cubic perovskite substrates with a lattice parameter of 3.868 Å. The lattice mismatch of VO$_2$ and LSAT is relatively small, -5.17% and 1.59% in $a$- and $c$-axis direction, respectively. Hence, LSAT can be a good candidate for epitaxial growth of high-quality VO$_2$ thin films.

In this work, we report our effort on growing heteroepitaxial VO$_2$ thin films on LSAT (111) substrates by PLD. We systematically investigated the film/substrate epitaxial relationships using high resolution X-ray diffraction. Raman spectroscopy studies provided us insights into the structural phase transitions in these thin films. We observed excellent electrical properties in these thin films with large four orders of magnitude change in resistivity between the M1 and R phase of VO$_2$.

## II. EXPERIMENTAL

### A. Thin Film Growth

Commercially available single crystal (111)-oriented LSAT wafers were used as substrates (purchased from MTI Corporation). The substrates were pretreated by annealing with a constant flow of O$_2$ at 1000°C for 4 hours. We used a 248 nm KrF excimer laser with an energy density of 1.5 J/cm$^2$ for the growth. Prior to the deposition, the chamber was evacuated to a background pressure of ~ 10$^{-7}$ Torr and then backfilled to an optimal oxygen partial pressure of 10 mTorr before heating up the substrate to the growth temperature of 500°C as measured by a thermocouple welded to the substrate heater. The VO$_2$ film was deposited from a dense polycrystalline target of V$_2$O$_5$. The target was sanded and fully pre-ablated to maintain a fresh surface for each growth. After the deposition, the samples were cooled at a rate of 5 °C/min to room temperature at an oxygen partial pressure of 10 mTorr.

### B. Characterization

Atomic force microscopy (AFM) was used to study the film surface morphology in the non-contact mode. The crystallinity of the films at room temperature (M1- structure) was identified by high resolution X-ray diffraction (XRD), which was performed using a triple-axis diffractometer equipped with a two bounce Ge (004) monochromator in a Bruker D8 Advanced diffraction system with Cu K$\alpha$ line ($\lambda$=0.154 nm) radiation. Raman spectroscopy measurements were performed in a backscattering geometry with a laser of 532 nm wavelength. The electrical properties of the $VO_2$ thin films were measured in the van der Pauw geometry. A rectangular sample (~5 mm × 2.5 mm) with ohmic contacts (15 nm Cr/200 nm Au) on the four corners of the stripe was prepared for the electrical measurements using thermal evaporation.

## III. RESULTS AND DISCUSSION

Fig. 1 (a) shows a representative XRD pattern from the out-of-plane $2\theta$-$\theta$ scan of the as-grown $VO_2$ thin film on LSAT substrate at room temperature. The well-resolved sharp reflections at 19.87°, 40.40°, 62.35° and 87.17° correspond to reflections from (111)-oriented LSAT substrate, namely LSAT 111, 222, 333, and 444, respectively. Only the $0k0$ family of reflections are visible for the $VO_2$ thin film over the range of 15°-95° in 2θ. This indicates that the $VO_2$ thin film is highly oriented along the out-of-plane direction of the LSAT substrate.

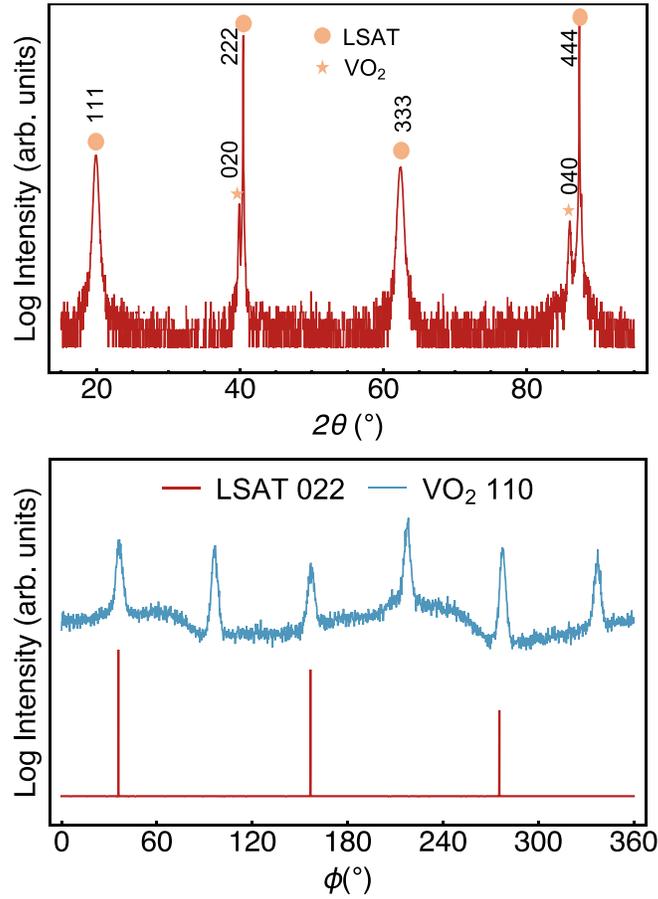

Fig. 1 (a). High resolution $2\theta$-$\theta$ XRD pattern of $VO_2$ thin film on LSAT substrate. (b). Off-axis $\varphi$ scan of $VO_2$ 110 and LSAT 022.

We used pole figure analysis to determine the in-plane film/substrate orientation relationships. We performed off-axis $\varphi$ scan for $VO_2$ 110 and LSAT 022 reflections. We observed six peaks separated by 60° for the $VO_2$ 110 reflection, whereas the LSAT 022 reflection showed three peaks separated by 120° (as shown in Fig. 1 (b)). Three of the peaks of $VO_2$ films were aligned with the substrate LSAT's peaks, whereas three other equally spaced peaks were found in between these matching peaks. This implies a direct one-to-one relationship between the film and substrate with a possible two types of in-plane domains for the $VO_2$ substrate. Based on these results, we conclude that $VO_2$ films in the M1 phase were epitaxially grown on LSAT (111) substrates with epitaxial relationships VO$_2$(020)||LSAT(111) and VO$_2$[001]||LSAT[11$\bar{2}$]. This result is consistent with previous reports on the growth of VO$_2$ thin films on STO (111) substrate.[10] Past investigations have established a correlation between the electrical properties of VO$_2$ and the structural domain sizes.[23] Hence, an understanding of the epitaxial relationship between the substrate and the film and the nature of VO$_2$ domains is crucial to achieve single crystal-like properties in thin film VO$_2$. In our case, we explored the structural evolution of the

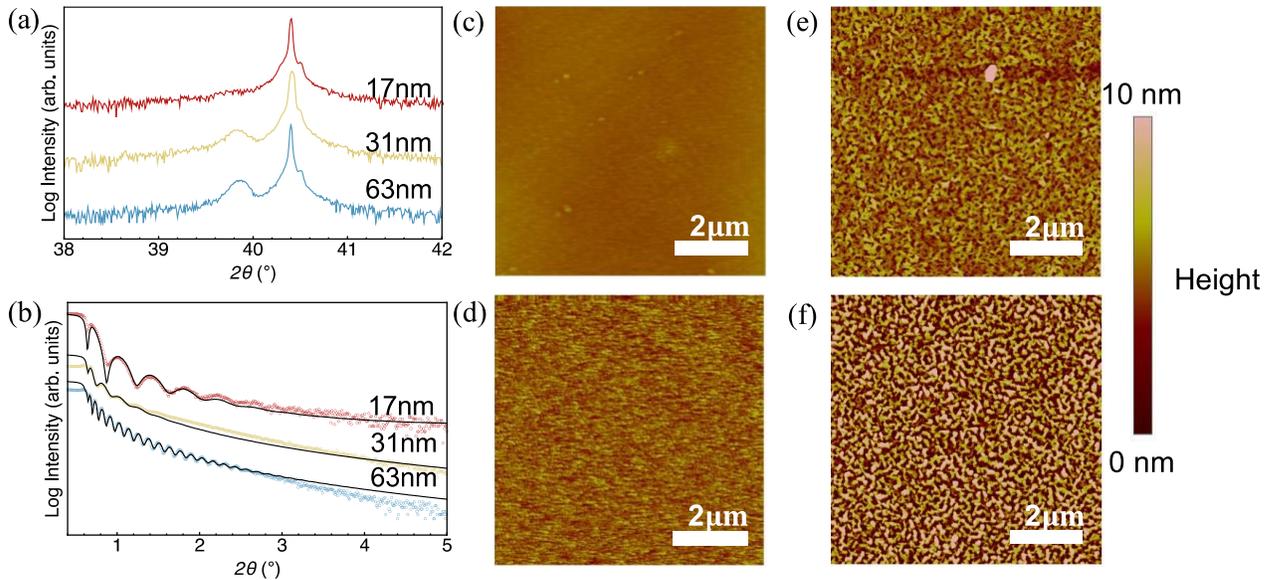

Fig. 2 Thickness dependent (a). 2$\theta$-$\theta$ XRD patterns and (b). X-ray reflectivity (XRR) of VO$_2$ thin film (open circle) and simulation curve fitting to XRR data (solid line). AFM topography image of (c). annealed LSAT surface and (d). 17 nm, (e). 31 nm, and (f). 63 nm as-grown VO$_2$ surface.

VO$_2$ films as a function of film thickness. The structural evolution, in terms of the texture, and surface roughness are summarized in Figure 2. The texture and crystallinity of these VO$_2$ films as characterized earlier using out-of-plane XRD is summarized in Fig. 2 (a). We studied the thickness dependence of surface roughness of the VO$_2$ films using X-ray reflectivity (XRR) analysis and AFM as shown in Figs. 2 (c-f). It is worth noting that films below 17 nm were hard to detect in out-of-plane XRD studies due to the small structure factor of $VO_2$ and the tiny volume of material that was probed. The presence of $VO_2$ films at these thicknesses was further confirmed by XRR. The thickness of the films as determined by the simulation of the measured XRR patterns were 17 nm, 31 nm and 63 nm, respectively. The slow decay of the thickness oscillations indicates that the surface of $VO_2$ thin films is smooth. This was further confirmed by AFM. The representative topography image of an annealed LSAT substrate is shown in Fig. 2 (c). We observed a relatively smooth surface with a root-mean-square (RMS) roughness of up to ~0.24 nm for the annealed LSAT (111) substrates. Please note that this value is larger than (001) oriented substrates due to the polar nature of the material and the complex chemical composition. In Fig. 2 (d), one can clearly see a relatively flat surface at 17 nm with an RMS roughness of ~1.1 nm. We observe the formation of sub-micron size islands as we increase the thicknesses to 31 and 63 nm as shown in Fig. 2 (e) and Fig. 2 (f) respectively. Meanwhile, the RMS roughness of the films increases to 3.3 nm and 5.1 nm for 31 nm- and 63 nm-thick film, respectively.

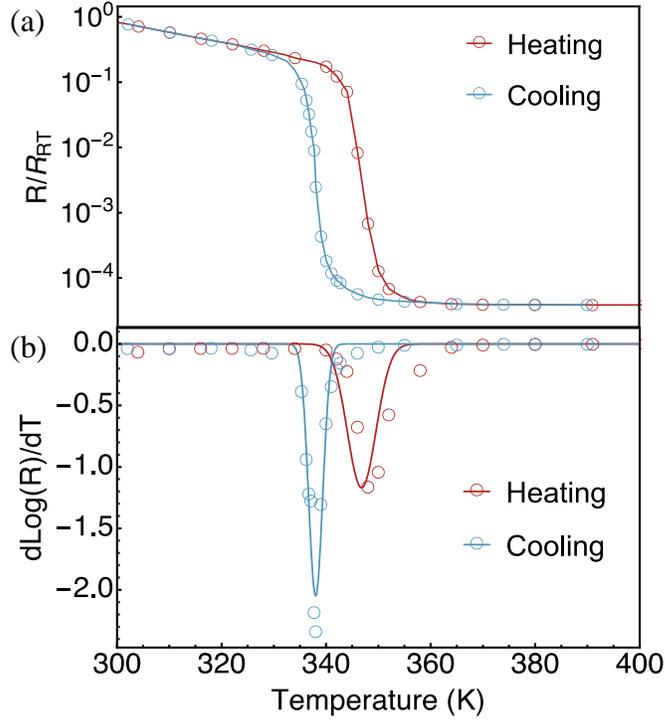

Fig. 3 (a) Temperature dependent four probe resistance of VO$_2$ grown on LSAT (111) substrate and (b) the corresponding derivative of the resistance. Lines in the derivative plot are Gaussian fits to the data.

Fig. 3 (a) shows the temperature dependence of the four-probe resistance of the VO$_2$/LSAT film from room temperature to 400 K for both the heating and cooling cycles. We see a large four orders of magnitude change in the resistance as we transform from M1 to R phase. The resistance ratio between the two phases, which can be defined as $R_{298K}/R_{380K}$, is ~ 3×10$^4$. This is among the largest resistance values observed for VO$_2$ grown on surfaces with 3*m* symmetry. The MIT transition temperature, defined as a peak position in a derivative curve of the R(T), is found to be ~346 K during heating as shown in Fig. 3 (b). This transition temperature is slightly larger than 340 K observed for bulk single crystals of VO$_2$. This small change could arise from the misfit strain imposed on the film by the substrate. The full-width-at-half-maximum (FWHM) of the derivative for the heating cycle is estimated to be 6 K, compared to single crystal data of about 0.1 K.[3] The broadening of the FWHM could be caused by defects in the film such as dislocations and multiple domains.

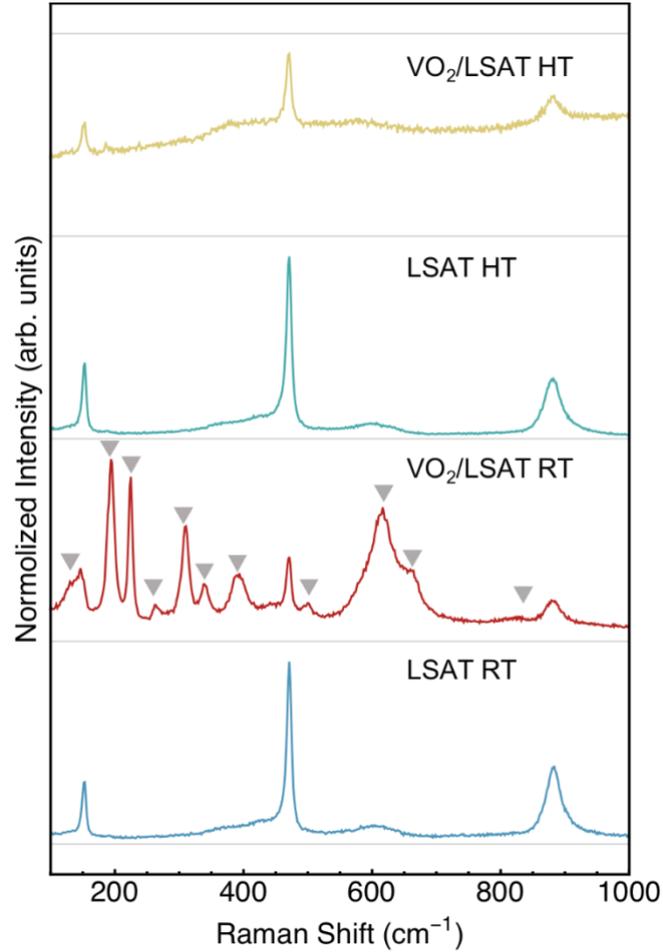

Fig. 4 Raman spectra of $VO_2$ thin film on LSAT substrate and LSAT substrate measured at 350 K (HT) and room temperature (RT) respectively. The 11 Raman active modes from room temperature $VO_2$ are highlighted by grey triangles. The zero baselines of the spectra are indicated by light grey lines.

To illustrate the correlations between the MIT and structural phase transition, we performed Raman spectroscopy study on $VO_2$ films on LSAT substrates and plain LSAT substrates at 298 K (RT) and 350 K (HT). The Raman spectra of LAST substrates at 350 K were nearly identical to that at room temperature. However, the $VO_2$/LSAT showed distinct Raman spectra for below and above transition temperature as shown in Fig. 4, presumably for the metallic and semiconducting phases, which have different structures. At room temperature, VO₂ has a distorted rutile structure with space group $C_5^{2h} = P2_1/c$ (14). Theoretically, there are nine $A_g$ and nine $B_g$ Raman active modes for this symmetry. In the Raman response of VO₂/LSAT at room temperature, one can clearly distinguish 11 modes for VO₂ M1 phase apart from the peaks attributed to LSAT substrates, corresponding to six $A_g$ and four $B_g$ modes. Detailed information about the nature of the Raman modes and the corresponding wave vectors are shown in Table 1. It is worth noting that we were unable to observe all the possible modes, possibly due to thermal smearing, as our measurements were carried out at room temperature, as compared to 85 K for the previous report.[24] The high temperature phase of VO₂ crystallizes in the rutile structure with a space group of $D_{14}^{4h} = P4_2/mnm$ (136). In the high temperature phase, four modes ($A_{1g}$, $B_{1g}$, $B_{2g}$, $E_g$) are Raman active. We observed a large background intensity enhancement in addition to the peaks from LSAT substrates in the Raman spectrum of VO₂/LSAT at 350 K. However, comparing to the background level in the Raman spectrum of LSAT substrates at 350 K, we can conclude that this background enhancement did not arise from any thermal effects. Notably, one important characteristic of the rutile TiO₂ Raman response is absence of strong, sharp peaks.[25] We believe that the broad enhancement in the background signal is due to the R phase of the VO₂ as observed in high-quality single crystals of VO₂.[26]

| This work (300K) | | Peter Schilbe[24] (85K) | |
|---|---|---|---|
| Raman Frequency (cm$^{-1}$) | Raman Mode | Raman Frequency (cm$^{-1}$) | Raman Mode |
| 146 | ... | 149 | ... |
| 195 | Ag | 199 | Ag |
| 224 | Ag | 225 | Ag |
|  |  | 259 | Bg |
| 263 | Bg | 265 | Bg |
| 310 | Ag | 313 | Ag |
| 340 | Bg | 339 | Bg |
| 391 | Ag | 392 | Ag |
|  |  | 395 | Bg |
|  |  | 444 | Bg |
|  |  | 453 | Bg |
|  |  | 489 | Bg |
| 500 | Ag | 503 | Ag |
|  |  | 595 | Ag |
| 613 | Ag | 618 | Ag |
| 662 | Bg | 670 | Bg |
| 827 | Bg | 830 | Bg |

Table. 1 Comparison of the observed Raman modes and the nature of these modes with the reference report.

## IV. SUMMARY AND CONCLUSIONS

In conclusion, we have presented high-quality epitaxial growth of $VO_2$ thin films on LSAT (111) substrates. The epitaxial relationship between the $VO_2$ thin film and the LSAT substrate was obtained by pole figure analysis. The pole figure patterns displayed the quasi-hexagonal azimuthal symmetry for $VO_2$, which resulted presumably from the two types of $VO_2$ domains on the surface with $3m$ symmetry. Raman spectroscopy demonstrated the concomitant structural phase transition of $VO_2$ thin film with the MIT. Finally, the $VO_2$ film revealed a MIT at 345 K with a large resistance ratio (~$3\times10^4$). These studies establish a pathway for future studies on the epitaxial growth of $VO_2$ thin films on substrates with $3m$ symmetry.

## ACKNOWLEDGMENTS

The authors gratefully acknowledge support from the Air Force Office of Scientific Research with grant no. FA9550-16-1-0335. S.N. acknowledges Link Foundation Energy Fellowship. The authors acknowledge the technical assistance of Shengyuan Bai and Boyang Zhao and use of Raman spectrometer of Prof. Steve Cronin at USC. The authors also acknowledge the Center for Excellence in Electron Microscopy and Microanalysis for the use of characterization studies.

[11] F.Y. Gan and P. Laou, J. Vac. Sci. Technol. A Vacuum, Surfaces, Film. **22**, 879 (2004).

[12] M.B. Sahana, G.N. Subbanna, and S. a. Shivashankar, J. Appl. Phys. **92**, 6495 (2002).

[13] M.-H. Lee and M.-G. Kim, Thin Solid Films **286**, 219 (1996).

[14] H. Jerominek and D. Vincent, Opt. Eng. **32**, 2092 (2015).

[15] C.O.F. Ba, V. Fortin, S.T. Bah, R. Vallée, and A. Pandurang, J. Vac. Sci. Technol. A Vacuum, Surfaces, Film. **34**, 031505 (2016).

[16] D.H. Kim and H.S. Kwok, Appl. Phys. Lett. **65**, 3188 (1994).

[17] D.-H. Youn, H.-T. Kim, B.-G. Chae, Y.-J. Hwang, J.-W. Lee, S.-L. Maeng, and K.-Y. Kang, J. Vac. Sci. Technol. A Vacuum, Surfaces, Film. **22**, 719 (2004).

[18] K. Kosuge, J. Phys. Chem. Solids **28**, 1613 (1967).

[19] Y. Cui and S. Ramanathan, J. Vac. Sci. Technol. A Vacuum, Surfaces, Film. **29**, 041502 (2011).

[20] M. Yang, Y. Yang, Bin Hong, L. Wang, K. Hu, Y. Dong, H. Xu, H. Huang, J. Zhao, H. Chen, L. Song, H. Ju, J. Zhu, J. Bao, X. Li, Y. Gu, T. Yang, X. Gao, Z. Luo, and C. Gao, Sci. Rep. **6**, 23119 (2016).

[21] L.L. Fan, Y.F. Wu, C. Si, G.Q. Pan, C.W. Zou, and Z.Y. Wu, Appl. Phys. Lett. **102**, 011604 (2013).

[22] F.J. Wong, Y. Zhou, and S. Ramanathan, J. Cryst. Growth **364**, 74 (2013).

[23] A. Moatti, R. Sachan, J. Prater, and J. Narayan, ACS Appl. Mater. Interfaces **9**, 24298 (2017).

[24] P. Schilbe, Phys. B Condens. Matter **316–317**, 600 (2002).

[25] Y. Zhang, C.X. Harris, P. Wallenmeyer, J. Murowchick, and X. Chen, J. Phys. Chem. C **117**, 24015 (2013).

[26] R. Srivastava and L.L. Chase, Phys. Rev. Lett. **27**, 727 (1971).
14